\documentstyle[12pt,amssymb]{article}

\def\beq{\begin{equation}} 
\def\eeq{\end{equation}}
\def\bea{\begin{eqnarray}}
\def\eea{\end{eqnarray}}
\def\bq{\begin{quote}}
\def\eq{\end{quote}}

\def\ASPM{{\it Advanced Studies in Pure Math.} }
\def\IJMP{{\it Int. J. Mod. Phys.} }
\def\JL{{\it JETP Lett.} }
\def\PL{{\it Phys. Lett.} }
\def\PR{{\it Phys. Rev.} }

\def\NP{{\it Nucl. Phys.} }

\begin{document}
\pagestyle{empty}

\begin{flushright}
RU-98-16\\
April 1998
\end{flushright}

\vspace*{1cm}

\begin{center}
{\Large\bf Non-Existence of Local Integrals of Motion} \\
\vspace*{0.2cm}
{\Large\bf in the Multi-Deformed Ising Model} \\
\vspace*{2.0cm}
{\large\sf Pedro D. Fonseca}\footnote{e-mail: {\sf
pfonseca@physics.rutgers.edu}}\\
\vspace*{0.5cm}
Department of Physics and Astronomy \\
Rutgers University \\
Piscataway, NJ 08855-0849 \\

\vspace*{1.5cm}
\end{center}
\begin{center}
{\bf Abstract}
\end{center}
We confirm the non-integrability of the multi-deformed Ising Model, an 
already expected result. After deforming with the energy operator
$\phi_{1,3}$ we use the Majorana free fermionic representation for the
massive theory to show that, besides the trivial one, no local
integrals of motion can be built in the theory arising from perturbing
with both energy and spin operators.

\vfill\eject

\setcounter{page}{1}
\pagestyle{plain}
%\vspace{0.5cm}

\section{Introduction}

After Zamolodchikov's work \cite{sasha1}-\cite{sasha3},
great advances have been achieved
in the understanding of field theories, and in particular of integrable field
theories (IFT), arising from the perturbation of certain conformal field
theories (CFT). The simplest example is certainly given by the perturbed 
Ising Model
\beq
{\cal A}={\cal A}_{CFT}+\tau\int\epsilon(z,\bar z)d^2z
+h\int\sigma(z,\bar z)d^2z~,
\label{eq:deformedaction}
\eeq
where $A_{\rm CFT}$ stands for the action of the two dimensional
$c=1/2$ CFT and $\sigma$ (spin) and $\epsilon$ (energy) are,
respectively, the relevant spinless primary fields $\phi_{1,2}$ and
$\phi_{1,3}$ with conformal dimensions $(1/16,1/16)$ and $(1/2,1/2)$.
From dimensional analysis, we see that the coupling constants
$h\sim\mbox{(length)}^{-15/8}$ and $\tau\sim\mbox{(length)}^{-1}$ have
conformal dimensions $(15/16,15/16)$ and $(1/2,1/2)$, respectively.

When considered separately, these perturbations were studied in
\cite{sasha1}-\cite{egushi}, and are
known to yield IFT's.  Namely, in the case $\tau\neq0$, $h=0$ the
corresponding perturbation has a realization as an Affine Toda Field
Theory (ATFT) based on $SU(2)$ (i.e. a sine-Gordon system) with an infinite
set of local integrals of motion (IM) of spin $s=1,3,5,7,9,\ldots$
(the Coxeter exponents modulo the dual Coxeter number of $SU(2)$),
while the case $\tau=0$, $h\neq0$ also has an infinite set of local IM
but with spin $s=1,7,11,13,17,19,23,29~(\mbox{mod}~30)$ (the Coxeter
exponents modulo the dual Coxeter number of $E_8$), related to the
fact that now it has a realization as an ATFT based on $E_8$.

Finally (the case that will interest us here), when both perturbations
are turned on (see \cite{dms} for an extensive analysis), 
although there is the possibility of having conserved
charges of spin $s=1,7,11,\ldots~(\mbox{mod}~30)$, in \cite{mussardo}
it is shown that no low spin IM exist, leading to belief that
(\ref{eq:deformedaction}) is no longer integrable, except for the above
$h=0$ or $\tau=0$ particular cases. In fact, a proof for it can be
obtained by starting with the finite $\tau$ theory (with scattering
matrix $S=-1$) and then, using perturbation theory in $h$ to compute the
corresponding $S$-matrix elements,
showing that there exists particle production for $h\neq0$
(see, for instance, \cite{mussardo} and \cite{dms}). 
In this letter we confirm this result by explicitly verifying the absence
of local IM, with the argument being as follows:

Given a minimal model, let  
$T_{s+1}$ be some descendant state of the identity operator with
spin $s+1$ satisfying the conservation law $\bar\partial T_{s+1}=0$ in
the CFT and $\phi_{kl}$ represent some perturbing relevant
spinless operator with operator product expansion (OPE)
\beq
T_{s+1}(z)\phi_{kl}(w,\bar w)\sim \ldots
+\frac{A_{kl}^{(s-1)}}{(z-w)^2}
+\frac{B_{kl}^{(s)}}{z-w}+\ldots~.
\eeq
From the deformed Ward identities we then have 
\beq
\bar\partial T_{s+1}=\lambda B_{kl}^{(s)}-\partial(\lambda
A_{kl}^{(s-1)})+\partial(\ldots)~,
\label{eq:barpartialT}
\eeq
where $\lambda$ is the perturbation coupling, so that
there exists a function $\Theta_{s-1}$ such that
\beq
\bar\partial T_{s+1}=\partial \Theta_{s-1}~,
\eeq
only if $B_{kl}^{(s)}$ can be written as a partial derivative
of some local field, 
i.e. $B_{kl}^{(s)}=\partial(\ldots)$. In such case, we can build the 
conserved charge of spin $s$
\beq
P_s=\int T_{s+1}dz+\Theta_{s-1}d\bar z~,
\label{eq:P}
\eeq
plus its anti-holomorphic partner, that we will not write down in what 
follows.
We will denote the set formed by $T_{s+1}$ in (\ref{eq:P}) 
for all possible values of $s$ by $\Lambda_{kl}$.
Whenever $\Lambda_{kl}$ has an infinite number 
of elements the $\phi_{kl}$ perturbed theory is said to be integrable.

In the case of the unitary minimal models deformed by 
$\phi_{1,3}$, $\Lambda_{1,3}$ can be derived from the sine-Gordon model, 
with the first elements given by \cite{sy}, \cite{egushi} 
\bea
T_2=T~,\indent 
T_4=(TT)~,\indent 
T_6=(T(TT))-\frac{(c+2)}{12}(\partial T\partial T)~,\indent
\ldots
\label{eq:T_s+1}
\eea
where $T(z)$ is the holomorphic part of the stress-energy tensor in the CFT
and the normal ordering $(AB)(w)$ is simply
\beq
(AB)(w)=\int\frac{dz}{2\pi i}\frac{A(z)B(w)}{z-w}~.
\eeq

In a minimal model deformed by two operators of conformal dimension
$h_1$ and $h_2$ and couplings $\lambda_1$ and $\lambda_2$, 
(\ref{eq:barpartialT}) is generalized to
\beq
\bar\partial T_{s+1}=\sum_{n,m}\lambda_1^n\lambda_2^m C^{(n,m)}+
\partial(\ldots)~,
\eeq
where $C^{(n,m)}$ are operators of conformal dimension
$(s-n(1-h_1)-m(1-h_2),1-n(1-h_1)-m(1-h_2))$, where the right dimension
can only take the allowed values for the particular model. In the
Ising Model case (\ref{eq:deformedaction}), these are
$\{0,\frac{1}{2},\frac{1}{16}\}$, implying that $n$ and $m$ cannot be
non-zero simultaneously and that only the linear terms on 
$\tau$ and $h$ survive (i.e. there are no resonance terms).

In conclusion, given $T_{s+1}\in\Lambda_{1,3}$ for the Ising model, we
will shown that, for $s\neq1$, 
$B_{1,2}^{(s)}$ appearing in the OPE with $\phi_{1,2}$
cannot be written as a partial derivative, thus
proving the non-existence of non-trivial local IM.

\section{Multi-Deformed Ising Model}

We start be writing an explicit form for $T_{s+1}\in\Lambda_{1,3}$ of
the $\phi_{1,3}$ perturbed Ising Model. For that we will use
its Majorana free massive fermion representation (with
$\phi_{1,3}=i(\psi\bar\psi))$,
\beq
S=\frac{1}{2\pi}\int d^2z\left(
\psi\bar\partial\psi
+\bar\psi\partial\bar\psi
+m\psi\bar\psi
\right)~,
\eeq
where $m\propto(T-T_C)$. This theory has $c=1/2$, OPE 
$\psi(z)\psi(w)\sim-1/(z-w)$ and stress-energy tensor 
given by $T(z)=-\frac{1}{2}(\psi\partial\psi)(z)$, where
$\psi(z)$ and $\bar\psi(\bar z)$ are primary fields 
with conformal weight $(1/2,0)$ and $(0,1/2)$ that we will
expand in modes as
\beq
\psi=\sum_n \frac{a_n}{z^{n+1/2}}~,
\label{eq:psi}
\eeq 
where $n\in{\Bbb Z}+\frac{1}{2}$ or $n\in{\Bbb Z}$ depending if we are
in the Neveu-Schwarz or in the Ramond sectors, respectively.
We get $\{a_n,a_m\}=\delta_{n+m,0}$, 
i.e. $a_n$ and $a_{-n}$ ($n>0$) are interpreted as being the usual 
annihilation and creation fermionic operators, and write 
the stress-energy tensor $T(z)=\sum_n L_n/z^{n+2}$ as
\beq
T(z)=\frac{1}{2}\sum_{n,m}\left(m+\frac{1}{2}\right)
\frac{:a_{n-m}a_m:}{z^{n+2}}~,
\label{eq:T}
\eeq
where $:\cdots:$ denotes the usual mode normal ordering.

Without loss of generality, $T_{s+1}\in{\Lambda_{1,3}}$ (in
(\ref{eq:T_s+1})) can be written as
\beq
T_{2k+2}(z)=(-1)^{k+1}\frac{1}{2}
(\partial^k\psi\partial^{k+1}\psi)(z)~,
\label{eq:T_2k+2}
\eeq
where we have set $s=2k+1$.

Since we are interested in the OPE of these quantities with $\sigma$
we will be working in the Ramond sector ($n,m\in{\Bbb Z}$ above), with
$\sigma(0,0)$ giving the Ramond vacuum
$|\frac{1}{16}\rangle=\sigma(0,0)|0\rangle$. Using (\ref{eq:psi}),
(\ref{eq:T_2k+2}) and the fact that $a_n\sigma=0$ for $n>0$ we get the
OPE $T_{2k+2}(z)\sigma(0,0)=\ldots+B^{(2k+1)}_{1,2}/z+\ldots$, where
\beq
B_{1,2}^{(2k+1)}=-\sum_{n=-2k-1}^{-k-1}c_n^{(k)}(a_na_{-2k-1-n}\sigma)~,
\label{eq:B}
\eeq
with coefficients
\beq
c_n^{(k)}=\left(n+\frac{1}{2}\right)
\left(n+\frac{3}{2}\right)\ldots\left(n+2k+\frac{1}{2}\right)~.
\eeq

Now, assume there exists an operator ${\cal O}^{(2k)}$ such that
\beq
B_{1,2}^{(2k+1)}=\partial {\cal O}^{(2k)}.
\label{eq:BpartialO}
\eeq
Obviously, ${\cal O}^{(2k)}$ 
is some level-$2k$ descendant operator of $\sigma$ and,
as will become clear soon, the best candidate can be written as
\beq
{\cal O}^{(2k)}=\sum_{n=-2k}^{-k}{d^{(k)}_n}a_na_{-2k-n}\sigma~,
\label{eq:O}
\eeq
where $d^{(k)}_n$ are some unknown coefficients and
again the limits simple arise from normal ordering and 
imposing $a_n\sigma=0$ for $n>0$. Taking the derivative of this operator
amounts to applying $L_{-1}$, which, from (\ref{eq:T}), can be written as
\beq
L_{-1}=-\sum_{m<-1/2}\left(m+\frac{1}{2}\right)a_ma_{-1-m}~.
\label{eq:L_-1}
\eeq

For $k=0$, (\ref{eq:B}) becomes
$B^{(0)}_{1,2}=\frac{1}{2}a_1a_0\sigma$ so that (\ref{eq:BpartialO})
is verified for $d^{(0)}_0=4$ in (\ref{eq:O}) (using (\ref{eq:L_-1})
$B^{(0)}_{1,2}$ is directly seen to be $L_{-1}\sigma$), implying that
indeed an $s=1$ IM exists.

For $k=1$, $B^{(3)}_{1,2}=\frac{3}{8}(5a_{-3}a_0-a_{-2}a_{-1})\sigma$,
while $L_{-1}{\cal O}^{(2)}=(\frac{5}{2}d^{(1)}_{-2}a_{-3}a_0-
\frac{1}{8}d^{(1)}_{-2}a_{-1}a_{-2})\sigma$, implying that
$d^{(1)}_{-2}$ has two satisfy two incompatible conditions and,
therefore, does not exist.  As to $k=2$,
$B^{(5)}_{1,2}=\frac{15}{32}(63a_{-5}a_0
-7a_{-4}a_{-1}+3a_{-3}a_{-2})\sigma$
and $L_{-1}{\cal O}^{(4)}=\frac{9}{2}d^{(2)}_{-4}a_{-5}a_0\sigma
+(\frac{7}{2}d_{-3}^{(2)}+\frac{1}{8}d_{-4}^{(2)})a_{-4}a_{-1}\sigma
+\frac{3}{2}d_{-3}^{(2)}a_{-3}a_{-2}\sigma$, so that (\ref{eq:BpartialO})
supplies three equations for $d^{(2)}_{-4}$ and
$d^{(2)}_{-3}$, again easily seen to be inconsistent.
This agrees with the absence of 
local IM of spin $s=3$ and $s=5$ in the Ising Model perturbed by a
magnetic field.

Finally, for $k\ge3$ some simple calculations yield
\bea
\partial {\cal O}^{(2k)}&=&L_{-1}{\cal O}^{(2k)}\nonumber\\
&=&-\sum_{n=-2k-1}^{-k-1}
e^{(k)}_na_na_{-2k-1-n}\sigma
+\frac{1}{2}\sum_{n=-2k+2}^{-k-1}d^{(k)}_n a_na_{-2k-n}a_{-1}a_0\sigma~,
\label{eq:partialO}
\eea
where
\bea
e^{(k)}_n&=&\left(n+\frac{1}{2}\right)d^{(k)}_{n+1}(1-\delta_{n,-k-1})
\nonumber\\
&&-\left(n+2k+\frac{1}{2}\right)d^{(k)}_n(1-\delta_{n,-2k-1})
(1-\delta_{n,-2k})
-\frac{1}{8}d^{(k)}_{-2k}\delta_{n,-2k}~.
\eea

The second term in (\ref{eq:partialO}) is obviously incompatible with
the desired structure in (\ref{eq:B}), implying that ${\cal O}^{(2k)}$
in (\ref{eq:BpartialO}) does not exist. Expressions other that the
quadratic chosen in (\ref{eq:O}) would obviously have more
incompatible terms, thus ending our proof.

\section*{Acknowledgments}

The author is most thankful to A. B. Zamolodchikov for having proposed
this problem, reviewed the manuscript and for the all the enlightening
and valuable conversations.  This work was supported by JNICT - PRAXIS
XXI (Portugal) under the grant BD 9102/96.


\newpage

\begin{thebibliography}{99}


\bibitem{sasha1} A. B. Zamolodchikov, \JL {\bf 46} (1987) 160.

\bibitem{sasha2} A. B. Zamolodchikov, \ASPM {\bf 19} (1989) 641.

\bibitem{sasha3} A. B. Zamolodchikov, \IJMP {\bf A4} (1989) 4235.

\bibitem{sy} R. Sasaki, I. Yamanaka, \ASPM {\bf 16} (1988) 271.

\bibitem{egushi} T. Egushi, S. K. Yang, \PL {\bf B224} (1989) 373.

\bibitem{mussardo} G. Mussardo, \PR {\bf 218}, 5\&6 (1992) 215.

\bibitem{dms} G. Delfino, G. Mussardo, P. Simonetti, 
\NP {\bf B473} (1996) 469.

\end{thebibliography}
\end{document}